\documentclass[aps,prc,twocolumn,showpacs,amsmath,preprintnumbers,amssymb,groupedaddress,superscriptaddress]{revtex4}
\usepackage{bbm}

\usepackage{graphicx}
\usepackage{dcolumn}
\usepackage{bm}
\usepackage{multirow}
\usepackage{hhline}

\usepackage{color}

\begin{document}

\preprint{}

\title{Directed flow of transported and non-transported protons in Au+Au collisions from UrQMD model}

\author{Yao Guo}

\author{Feng Liu}
\thanks{fliu@iopp.ccnu.edu.cn}
\affiliation{Institute of Particle Physics, Central China Normal University, WuHan,HuBei 430079, China}
\affiliation{The Key Laboratory of Quark and Lepton Physics (Central China Normal University), Ministry of Education, Wuhan, Hubei, 430079, China}
\author{Aihong Tang}

\affiliation{Physics Department, Brookhaven National Laboratory, Upton, New York 11973, USA}

\date{\today}

\begin{abstract}
The directed flow of inclusive, transported and non-transported (including produced) protons, as well as antiprotons, has been studied in the framework of Ultra-Relativistic Quantum Molecular Dynamics approach (UrQMD model) for Au+Au collisions at $\sqrt{s_{NN}}=$7.7, 11.5, 19.6, 27, 39, 62.4 and 200 GeV. The rapidity, centrality and energy dependence of directed flow for various proton groups are presented. It is found that the integrated directed flow decreases monotonically as a function of collision energy for $\sqrt{s_{NN}}=$11.5 GeV and beyond. However, the sign-change of directed flow of inclusive protons, seen in experimental data as a function of centrality and collision energy, can be explained by the competing effect of directed flow between transported and non-transported protons. Similarly the difference in directed flow between protons and antiprotons can be explained. Our study offers a conventional explanation on the cause of the $v_1$ sign-change other than the antiflow component of protons alone which is argued to be linked to a phase transition.
\end{abstract}


\maketitle

\section{Introduction}

Relativistic Heavy Ion collisions, involving two nuclei moving close to the speed of light, produce
thousands of particles. The collective motion of the outgoing particles
is of special interest because it carries important information about the bulk properties of the sytem.
A handle into the study of such collective motion makes use of the azimuthal distributions of detected particles in non-central collisions~\cite{flowReview}. For a quantitative description, such distribution is conventionally decomposed~\cite{MethodPaper} by a Fourier series
\begin{equation}
E\frac{d^{3}N}{d^{3}p}=\frac{1}{2\pi}\frac{d^{2}N}{p_{T}dp_{T}dy}(1+\sum_{n=1}^{\infty}2v_{n}\cos
n\phi)
\end{equation}
where $\phi$ denotes the angle between the particle's azimuthal angle in momentum space and the reaction plane angle. The reaction plane is defined by the collision axis and the line connecting the centers of two nuclei. The various coefficients in this expansion can be calculated as~\cite{MethodPaper}:
\begin{equation}
v_{n}=\langle\cos n\phi\rangle
\end{equation}
The first and the second coefficients are refered to  as directed $(v_{1})$ and elliptic flow $(v_{2})$, respectively, and they both play important roles in describing the collective expansion in azimuthal space. Elliptic flow is produced by the conversion of the initial coordinate-space anisotropy into the momentum-space anisotropy, due to the developed large in-plane pressure gradient~\cite{Ollitrault92,Voloshin96,Heinz00}. Directed flow, which is the focus of this study, describes the ``side splash" of particles measured off mid-rapidity\cite{Sorge}. It probes the dynamics of the system in the longitudinal direction. The shape and magnitude of directed flow in the vicinity of midrapidity are of special interest, specially for protons,  because they are sensitive to the equation of state (EOS) and may carry phase transition signal\cite{antiFlow, thirdFlow,Stocker}.

Directed flow has been measured at the AGS~\cite{e895KShort,e895Lambda}, at the CERN SPS~\cite{na49}, and at RHIC~\cite{v1v4,phobosV1,v1At62GeV,v1SysSizeIndependent,PIDv1,lowEnergyV1}. In particular,
the directed flow of protons at RHIC has been studied in detail by the STAR Collaboration, for Au+Au collisions at $\sqrt{s_{NN}}=200$ GeV~\cite{PIDv1} and at lower energies~\cite{lowEnergyV1}.
 It is worthy of notice, that at energies studied by STAR, protons and antiprotons exhibit different $v_1(y)$ slope in close-to-central collisions (5-30\% at 200 GeV and 10-40\% at low energies), and at low energies the proton $v_1(y)$ slope changes its sign from negative as in peripheral collisions to positive as in close-to-central collisions. The sign-change of $v_1(y)$ slope has been
 proposed as a possible signal of phase transition, it is thus  necessary to identify other causes that may give rise to the similar sign-change of directed flow in the vicinity of y=0. Indeed it is mentioned in ~\cite{PIDv1} that the observed $v_1$ for protons is a convolution of the directed flow of produced protons with that of protons transported from beam rapidity. Since transported protons come from the original projectile and target nuclei, they preserve the directed flow of spectators. Produced protons, on the other hand, tend to flow with the bulk. The flow direction of spectators and bulk are not necessarily equal;
 the final $v_1$ observed would then be the result of a
competition between the two different proton sources.

In this work we study the interplay between the directed flow of transported protons and that of non-transported protons, within the framework of the Ultra-relativistic Quantum Molecular Dynamics model (UrQMD)~\cite{UrQMD}. Note that although $v_1(y)$ calculated with UrQMD does not reproduce the measurement from Au+Au collisions at $\sqrt{s_{NN}}=200$ GeV, as already pointed out in~\cite{PIDv1}, the study of the
contributions from transported and non-transported protons to the overall proton $v_1$
can offer insights into the competing dynamics which lead to the final observation of the sign of $v_1(y)$ and
its slope in the vicinity of midrapidity.

\section{Transported and non-transported particles in the UrQMD model}
UrQMD is a microscopic transport model based on the covariant propagation of all hadrons along classical
trajectories in combination with stochastic binary scatterings, color string formation and resonance decay. This microscopic transport model describes the phenomenology of hadronic interactions at low and intermediate energies ($\sqrt{s} < 5$ GeV) in terms of interactions between known hadrons and their resonances. At higher energies($\sqrt{s} > 5$ GeV), the excitation of color strings and their subsequent fragmentation into hadrons dominates the multiple production of particles in the UrQMD model~\cite{UrQMD}.
\begin{figure}
\centering
\includegraphics[height=13pc,width=20pc]{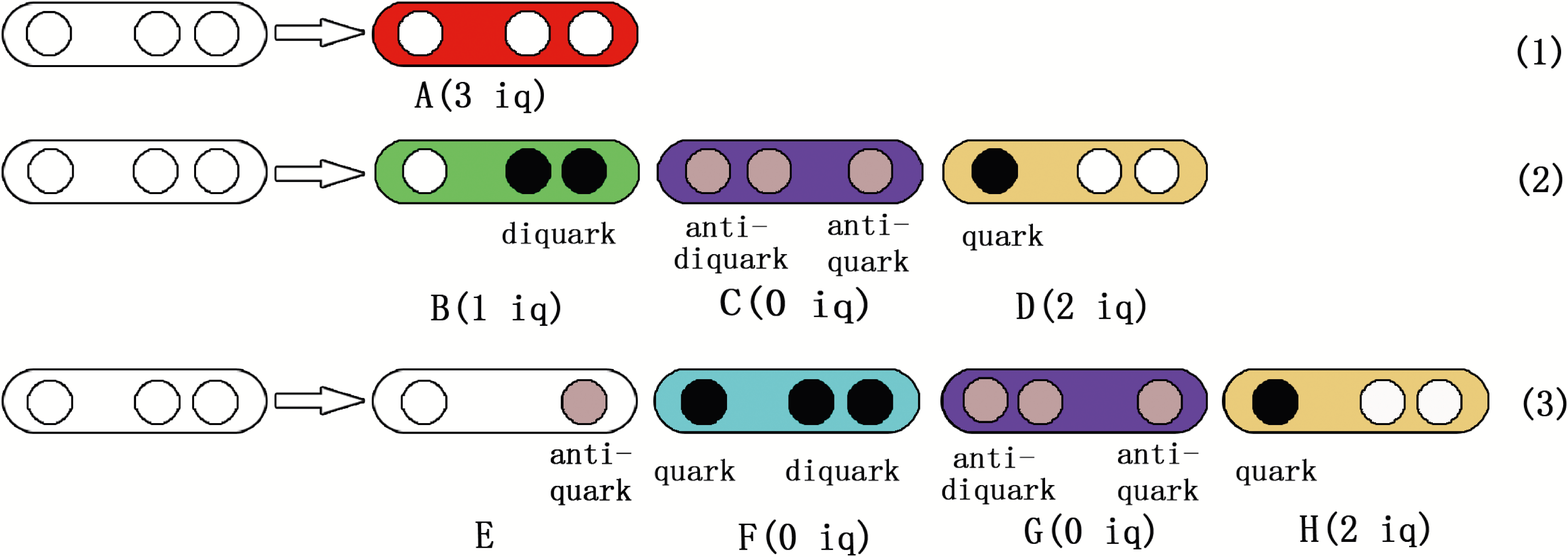}
\caption{Particle production in a string-excitation scheme in UrQMD. White circles, black circles and gray circles stand for initial quarks, produced quarks and produced antiquarks, respectively. In case (1), no quark pair is created. In case (2), one diquark-antidiquark pair and one quark-antiquark pair are spontaneously created in the color flux-tube between the initial quark and the initial diquark.  In case (3), similarly, two quark-antiquark pairs and one diquark-antidiquark pair are created, and one meson is produced.} \label{cartoon}
\end{figure}

Fig.~\ref{cartoon} schematically sketches the exitation and subsequent fragmentation ($\sqrt{s} > 5$ GeV) of
a baryon-string \cite{UrQMD}: In the first case, shown on top of the figure, the exitation energy is too low
to produce new quark pair(s), and there is only one baryon A in the final state. In the second case, one
diquark-antidiquark pair and one quark-antiquark pair are created. One of the initial quarks (white circle)
combines with a newly produced diquark (black circles) to form the baryon B, the newly produced anti-diquark
(brown circles) combines with a newly produced anti-quark to form the antibaryon C, and the newly produced
quark together with the initial diquark form another baryon D. The third case is similar to the second one,
and it shows the produced baryons and a meson. At the end baryon A has three initial quarks (3 iq,
where  iq stands for ``initial quark"), baryon D and H each have two initial quarks (2 iq), baryon B has one
initial quark (1 iq), and baryon G has zero initial quark (0 iq). The two anti-baryons (C and G) each
 have zero initial quark (0, iq). The more initial quarks can be found in a particle, the more initial
information that particle carries. By tracing the number of initial quarks in a particle, we classify
particles as produced or transported particles.  In this article, baryons with three initial quarks are
regarded as transported baryons, and all other baryons that have at least one produced quark are regarded
as non-transported. A particular subgroup of the latter are the produced baryons with zero initial quarks.
Produced baryons should be similar to anti-baryons in many respects as both groups are made of produced
quarks.

After the process of fragmentation excitation of baryon-strings, baryons produced (mostly unstable baryons) are subject to multiple scatterings. When the energy of a binary collision is lower than 5 GeV ($\sqrt{s} < 5$ GeV), there will be no strings involved and there are no new quark pairs produced. During this process the unstable baryons do decay but the number of constituent
quarks does not change. For example, in the case of a $\Lambda$ decay into a proton and a pion, one can still identify how
many quarks in the daughter proton are originated from initial quarks by tracing the quark constituent of its parent, $\Lambda$.

\section{Analysis and results}

\begin{figure}
\centering
\includegraphics[height=15pc,width=20pc]{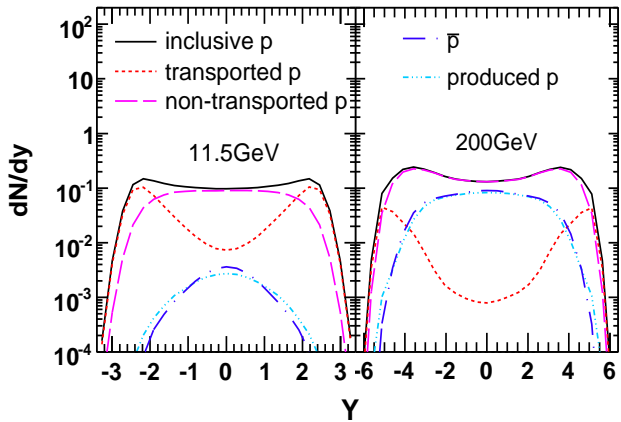}
\caption{UrQMD simulation of rapidity distribution of inclusive, transported, non-transported, produced, and anti-protons. The simulation is made for 10\%-70\% central Au+Au collisions collisions at $\sqrt{s_{NN}}=11.5$ GeV (left) and 200 (right) GeV.} \label{yield}
\end{figure}

Fig.~\ref{yield} shows the rapidity distribution of protons and anti-protons for Au+Au collisions at $\sqrt{s_{NN}}=11.5$ GeV and 200 GeV, calculated with the UrQMD model. The selection on centrality (10\%-70\%) and transverse momentum $p_T$ ($0.4-1$ GeV/$c$) are chosen to match those used in ~\cite{PIDv1}. The distribution of transported, non-transported, and produced protons are drawn separately.  As expected, the transported protons peak at beam rapidity and the produced ones, at midrapidity. Not
surprisingly, the distribution of produced protons is similar to that of anti-protons (to first order), and the remaining
difference comes from their separate production mechanisms . For example, for each produced baryon there is an accompanying
anti-baryon produced but that statement is not necessarily true for anti-baryons. In addition, antiprotons are subject to
annihilation while protons are not.  The transported (produced) protons accounts for 12\% (2.1\%) of inclusive protons at
11.5 GeV, and 0.91\% (50\%) at 200 GeV. The baryon stopping effect is seen at both energies. At midrapidity, transported
protons dominate over produced protons at 11.5 GeV, while at 200 GeV, the opposite is true. Note that the sum of the yield of
transported and produced protons does not give the inclusive proton yield because some of the protons can be regarded neither
as transported nor produced. Those are protons that have 2 or 1 initial quarks, and they account for a large fraction of the
total number of protons. By definition the inclusive proton yield is a sum of transported and non-transported proton yields.

\begin{figure}
\centering
\includegraphics[height=15pc,width=20pc]{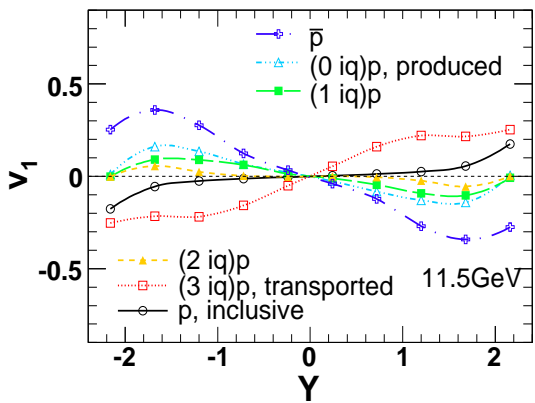}
\caption{Rapidity dependence of $v_{1}$ for protons and anti-protons in Au+Au collisions at $\sqrt{s_{NN}}=11.5$ GeV for centrality 10\%-70\%, from UrQMD calculation. See text for explanation on individual protons groups.}
\label{v1_y}
\end{figure}

To illustrate the effect of the mixture of initial and produced quarks on the shape of $v_1(y)$ Figure ~\ref{v1_y}
shows $v_1(y)$  plotted for all proton (antiproton) groups for Au+Au collisions collisions at $\sqrt{s_{NN}}=11.5$ GeV. It is
seen that for transported protons (3 iq), the $v_1(y)$ slope is positive at midrapidity. With the replacement of even only
one initial quark by a produced quark, as shown by the case (2 iq), the $v_1(y)$ shape becomes extremely flat at midrapidity.
As more initial quarks are replaced, the $v_1(y)$ slope changes sign to negative and continues to decrease monotonically with
descreasing number of initial quarks.  For produced protons (0 iq), the sign of the $v_1(y)$ slope agrees with that of
antiprotons as expected because both are produced particles. The $v_1(y)$ of inclusive protons is a convolution of
$v_1(y)$ and the relative yields of each individual proton group. For clarity, the figures shown below do not display the two intermediate proton groups (1 iq and 2 iq groups).

\begin{figure}
\centering
\includegraphics[height=14pc,width=20pc]{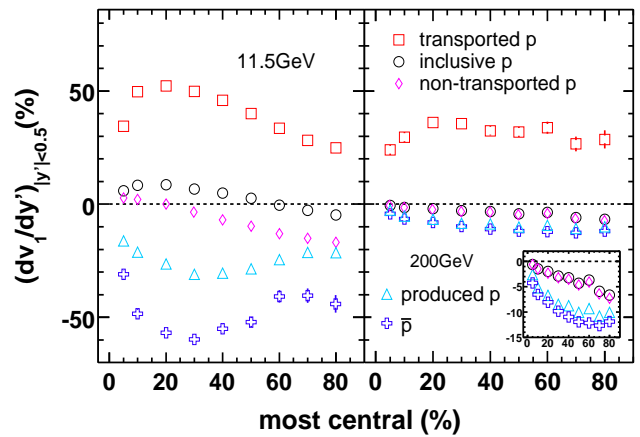}
\caption{The $v_{1}$ slope as a function of centrality for protons and anti-protons for Au + Au collisions at $\sqrt{s_{NN}}=5$ 11.5 (left) and 200 (right) GeV, calculated with the UrQMD model.} \label{v1_cent}
\end{figure}

In Fig.~\ref{v1_cent} we present the centrality dependence of $v_1(y')$ slope at midrapidity in Au+Au collisions at $\sqrt{s_{NN}}=11.5$ GeV and 200 GeV, for transported, non-transported, produced, and inclusive protons. The slope for antiprotons is also
shown for comparison. Here the slope is extracted from the normalized ($y^{'}=y/y_{beam}$) rapidity distribution, where
$y_{beam}$ is the beam rapidity. Our results are obtained by integrating over transverse momentum ($p_{T}$) from 0.4 to
1.0 GeV as done in ~\cite{PIDv1}.  At 11.5 GeV, as the overlap between colliding ions range from central to peripheral,
the $v_1(y')$ slope of transported protons increases rapidly, reaches a maximum at $\sim$20\% centrality and then decreases.
The opposite trend is seen for produced protons. The slope of $v_1(y')$ extracted from non-transported protons is negative in
most centralities, ranges between positive and negative values, as non-transported protons also include intermediate proton
groups other than produced protons. The $v_1(y')$ slope of inclusive protons is the result of competing effects from  directed
flow between transported and non-transported protons. Because the yield of non-transported protons dominate over that of
transported protons, in peripheral collisions the sign of the $v_1(y')$ slope agree with that of non-transported protons.
As the centrality decreases, transported proton $v_1(y')$ slope increases. This effect gradually overcomes the effect of
directed flow from non-transported protons and change the sign of the $v_1(y')$ slope to positive. In most central collisions,
$v_1$ from all groups vanishes due to the symmetrical shape of the collision. This explains the sign change of $v_1(y')$ seen
in data~\cite{lowEnergyV1}. Similarly, the trend of the $v_1(y')$ slope at 200 GeV can be explained with similar arguments.
Note that at 200 GeV, the $v_1(y')$ slopes are negative at all centralities. This can be due to the fact that at 200 GeV the
transported proton yield is too small and the overall effect is driven by non-transported protons. For the same reason, a
difference between proton and antiproton $v_1$ is observed as seen in the inset, similar to what is shown in ~\cite{PIDv1}.

\begin{figure}
\centering
\includegraphics[height=15pc,width=20pc]{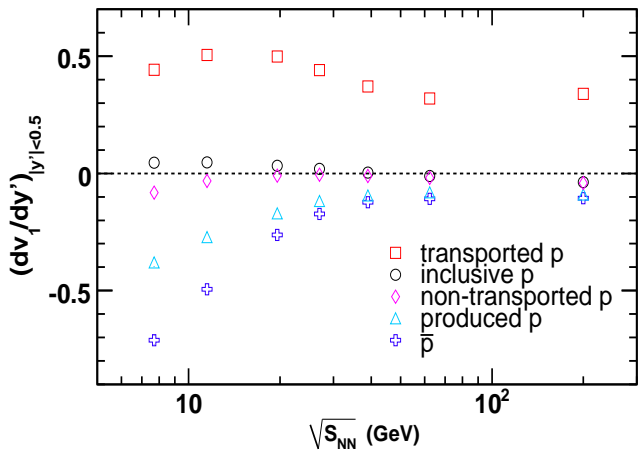}
\caption{UrQMD calculation of protons' and anti-protons' $v_1$ slope (d$v_{1}/dy^{'}$) around mid-rapidity ($|y^{'}|<0.5$) as a function of incident energy. } \label{energyDependence}
\end{figure}

The energy dependence of directed flow for protons and anti-protons is presented in Fig.~\ref{energyDependence}. Based on
UrQMD calculation, the slopes of $v_1(y')$ extracted from produced protons are negative and they follow more or less a similar
behavior as the one extracted from
 antiprotons, while the $v_1(y')$ slopes for transported protons are all positive. The $v_1(y')$ slope for inclusive protons
is a convoluted effect produced by transported and non-transported protons. It is positive at 7.7, 11.5,19.6, 27 and 39 GeV,
decreases monotonically with increasing energy for $\sqrt{s_{NN}}=$11.5 GeV and beyond, and becomes negative at 62.4 GeV.
The sign change of $v_1(y')$ slope is also seen in experimental data~\cite{lowEnergyV1}, although the energy at which the
$v_1(y')$ slope changes  sign is lower (11.5 GeV). These findings suggest that the apparent ``collapse" of $v_1$, defined by a
flat shape of the proton $v_1(y')$,  or the sign-change of the proton $v_1(y')$ slope, can be explained by the interplay
between transported protons and non-transported protons.

\section{Summary}
Directed flow of inclusive, transported protons and non-transported protons (including produced protons), as well as antiprotons,
 have been studied with the UrQMD model at RHIC energies. It is found that the integrated directed flow decreases monotonically
as a function of collision energy for $\sqrt{s_{NN}}=$11.5 GeV and beyond. However, the sign-change of directed flow of
inclusive protons, as a function of centrality and collision energy, can be explained by the competing effect of directed flow
between transported and non-transported protons. Similarly, the difference in directed flow between protons and antiprotons can
be explained. Our study identified alternative cause of the $v_1$ sign-change other than the anti-flow component of protons
alone which is argued to be linked to a phase transition.

\section{Acknowledgement}

This work is supported in part by National Natural Science Foundation of China under Grants 11075060 and 11135011. A. Tang is supported by U.S. Department of Energy under Grants DE- AC02-98CH10886 and DE-FG02- 89ER40531. We wish to thank G. Wang, D. Keane and J. Y. Chen for their valuable comments and suggestions. We thank R. Debbe for English corrections.


\begin{thebibliography}{99}
\bibitem{flowReview}
For a recent review, see : S. Voloshin, A. Poskanzer and R. Snellings, Volume 23, In {\it Relativistic Heavy Ion Physics}, Published by Springer-Verlag. Edited by R. Stock. DOI: 10.1007/978-3-642-01539-7. arXiv:0809.2949.

\bibitem{MethodPaper}
A.M.Poskanzer and S.A. Voloshin, Phys.Rev.C \textbf{58},1671 (1998).

\bibitem{Ollitrault92}
J.-Y. Ollitrault, Phys. Rev. D \textbf{46}, 229 (1992).

\bibitem{Voloshin96}
S. Voloshin and Y. Zhang, Z. Phys. C \textbf{70}, 665 (1996).

\bibitem{Heinz00}
P. Kolb, J. Sollfrank, and U. Heinz, Phys. Rev. C \textbf{62}, 054909 (2000).

\bibitem{Sorge}
H.Sorge, Phys. Rev. Lett. \textbf{78},2309 (1997) .

\bibitem{antiFlow}
J. Brachmann {\it et al.},  Phys. Rev. C {\bf 61}, 024909
(2000).

\bibitem{thirdFlow}
L. P. Csernai and D. R\"{o}hrich,  Phys. Lett. B {\bf 458},
454 (1999).

\bibitem{Stocker}
H.Stocker, Nucl. Phys. A \textbf{750} ,121(2005) .

\bibitem{e895KShort}
P. Chung {\it et al.} (E895 Collaboration), Phys. Rev. Lett.
{\bf 85}, 940 (2000).

\bibitem{e895Lambda}
P. Chung {\it et al.} (E895 Collaboration),  Phys. Rev. Lett.
{\bf 86}, 2533 (2001).

\bibitem{na49}
C. Alt {\it et al.} (NA49 Collaboration), Phys. Rev. C {\bf
68}, 034903 (2003).


\bibitem{v1v4}
J. Adams {\it et al.} (STAR Collaboration), Phys. Rev. Lett. {\bf 92}, 062301 (2004).

\bibitem{phobosV1}
B. B. Back {\it et al.} (PHOBOS Collaboration),  Phys. Rev. Lett {\bf 97}, 012301 (2006).

\bibitem{v1At62GeV}
J. Adams {\it et al.} (STAR Collaboration),  Phys. Rev. C {\bf 73}, 034903 (2006).

\bibitem{v1SysSizeIndependent}
B. I. Abelev {\it et al.} (STAR Collaboration),  Phys. Rev. Lett. {\bf 101}, 252301 (2008).

\bibitem{PIDv1}
L. Adamczyk {\it et al.} (STAR Collaboration), Phys. Rev. Lett. {\bf 108} 202301 (2012).

\bibitem{lowEnergyV1}
Y. Pandit (for the STAR Collaboration), arXiv:1112.0842;

\bibitem{UrQMD}
S. A. Bass, M. Belkacem, M. Bleicher, M. Brandstetter, L. Bravina, C. Ernst, L. Gerland, M. Hofmann, S. Hofmann, J. Konopka, G. Mao, L. Neise, S. Soff, C. Spieles, H. Weber, L. A. Winckelmann, H. Stocker, W. Greiner, Ch. Hartnack, J. Aichelin and N. Amelin, Prog. Part. Nucl. Phys. \textbf{41}, 255 (1998). \\
M. Bleicher, E. Zabrodin, C. Spieles, S.A. Bass, C. Ernst, S. Soff, L. Bravina, M. Belkacem, H. Weber, H. Stocker, W. Greiner, J. Phys. G. Nucl. Part. Phys. \textbf{25}, 1859 (1999).










\end{thebibliography}
\end{document}